\documentclass[11pt]{article}
\usepackage{fullpage,amsthm}
\usepackage{amsmath,amssymb,amsfonts}
\usepackage{algorithm}
\usepackage{algpseudocode}
\usepackage{graphicx}
\usepackage{textcomp}
\usepackage{xcolor}
\usepackage{lineno}
\usepackage{longtable}
\usepackage{indentfirst}
\usepackage{subcaption}
\usepackage{tabularx}
\usepackage{booktabs}
\usepackage{hyperref}
\usepackage{makecell}

\DeclareMathOperator*{\argmin}{arg\,min}

\begin{document}
\baselineskip 12pt

\begin{center}
\textbf{\Large \Large Carruthers Data Processing Pipeline: \\ Photon Background Removal} \\

\vspace{1.5cc}
{ \sc Alex Zhang${^{*1}}$, Heather Filippini${^2}$, Jackson Craig${^1}$, Lara Waldrop${^1}$, John Clarke${^3}$, Farzad Kamalabadi${^1}$, Pratik Joshi${^4}$}\\

\vspace{0.3 cm}

{\small $^{1}$Department of Electrical and Computer Engineering, University of Illinois at Urbana-Champaign \\
$^{2}$Illinois Coordinated Sciences Laboratory, University of Illinois at Urbana-Champaign 
\\
$^{3}$ Department of Astronomy and Center for Space Physics, Boston University \\
$^{4}$ Laboratory for Atmospheric and Space Physics, University of Colorado at Boulder \\
}
 \vspace{0.3 cm}
{\small $^{*}$Corresponding Author: alexmz2@illinois.edu}
 \end{center}
\vspace{1.5cc}

\begin{abstract}
  \noindent  The Carruthers Geocorona Observatory, launched in September 2025, is NASA’s first mission devoted to investigating the fundamental nature of Earth’s exosphere from its distant vantage in halo orbit around the Earth-Sun Lagrange 1 (L1) point. Its primary payload, the GeoCoronal Imager, consists of two coaligned photometric imagers that measure ultraviolet Lyman-$\alpha$ emission radiance from exospheric hydrogen simultaneously at wide- and narrow- fields of view. These observations will map the exosphere’s global spatial structure and observe its temporal variability in response to geomagnetic storms. However, a critical step in that analysis is isolating the in-band exospheric hydrogen Lyman-alpha signal from any other source of photons, including in-band InterPlanetary Hydrogen photon background and out-of-band photon backgrounds, which are emitted from Earth's limb. This paper details the algorithms used to retrieve and remove photon backgrounds from the GCI science images acquired on-orbit. Finally, the science data processing pipeline that transforms instrument-effect corrected images (L1B science data product) into absolutely-calibrated exospheric H measurements in physical units (L1C science data product) is detailed. Evaluation of algorithm performance based on a realistic pre-flight case study using synthetic data demonstrates that these photon background removal algorithms achieve high accuracy, leaving a residual systematic bias in isolated exospheric Lyman-$\alpha$ of only $3\%$ under beginning-of-life conditions.
\vspace{0.95cc}

\parbox{24cc}{{\it Key words and Phrases}: Earth's exosphere, Geocorona, Carruthers Geocorona Observatory, GCI instrument, Calibration
}
\end{abstract}

\section{Introduction}

The Carruthers Geocorona Observatory is a NASA Heliophysics Mission of Opportunity, which launched in September, 2025, with the goal of revealing the structure and dynamics of Earth's vast outermost atmospheric layer known as the exosphere.  From its distant vantage in halo orbit around the Earth-Sun L1 Lagrange equilibrium point, the Carruthers GeoCoronal Imager (GCI) will acquire near-continuous, high resolution, and high cadence images of exospheric airglow emission on global scales, revealing the exosphere's climatological variation and its transient response to geomagnetic storms for the first time.  The GCI targets the bright Lyman-$\alpha$ line at 121.6 nm, which is produced from resonant scattering of solar Lyman-$\alpha$ photons by the exosphere's constituent hydrogen (H) atoms. Calibrated GCI images of the exospheric Lyman-$\alpha$ radiance distribution are used to constrain retrievals of the underlying exospheric H density distribution, Carruthers' primary science data product, which in turn are used to address important science questions about the nature and origin of exospheric variability \cite{Waldrop26}.

The success of the Carruthers mission rests critically on the accuracy of GCI image calibration, which is performed in three major stages, each culminating in a formally identified Level Data Product as summarized in Table \ref{tab:data_product_def_table}.  The data processing pipeline begins with the raw data telemetered from the spacecraft, which are archived as Level 0 (L0) files.  The first two calibration steps, namely from L0 to Level 1A (L1A) and from L1A to Level 1B (L1B), result in the correction of all instrument effects, including image registration, flat fielding, and instrument background subtraction (see Zhang et al. \cite{Zhang26a} for details).  This paper is focused on the final calibration pipeline step, which converts L1B images of Earth's nadir into Level 1C (L1C) data by isolating the exospheric Lyman-$\alpha$ signal from sources of photon background and converting the resulting signal into physical units of emission radiance. L1C images are the foundational input to the H density retrieval algorithms used to produce Level 2 (L2) science data and address the scientific objectives of the Carruthers mission.  Because photon backgrounds are generally much more significant relative to the target exospheric signal compared to instrument backgrounds, the accuracy of their removal in L1C data has a correspondingly larger influence on the accuracy of the retrieved H density distributions.  

\begin{center}
\begin{longtable}{|c|c|c|} 
\caption{List of Data Product Definitions} \label{tab:data_product_def_table}
 \\
\hline
Data Product & Description & Units \\
\hline
L0 & Raw images from spacecraft & [DN] \\
L1A & Dark rows and bias frame removed & [DN] \\
L1B & All instrument effects removed and dark corners zeroed out & [DN] \\
L1C & All photon backgrounds removed, cross-calibrated & $\left[\frac{\text{ photons}}{\text{cm}^2 \cdot \text{s}} \right]$ \\
L2 & 3D Hydrogen density & $\left[\frac{\text{atoms}}{\text{cm}^3} \right]$ \\
\hline
\end{longtable}
\end{center}

Several photon background sources are present in the Earth nadir scene and must be detected and removed prior to scientific image analysis.  Stars, the Moon, and other solar system objects, which appear in the scene intermittently, are generally masked and precluded from use in science analysis as described below in Section \ref{sec:celestial_body_alg}.  The estimation of background Lyman-$\alpha$ emission, which is produced from resonant solar photon scattering by InterPlanetary Hydrogen (IPH) atoms distinct from those populating Earth's exosphere, is described in Section \ref{sec:iph_photon_retrieval_algorithm}. The IPH scene also supports cross-calibration between the various GCI imaging configurations, as described in Section \ref{sec:cross_channel_responsivity}. While the GCI is designed to target the bright H Lyman-$\alpha$ emission line, its suppression of non-Lyman-$\alpha$ (out-of-band, or OOB) photons is incomplete.  Estimation and removal of OOB photon contamination is generally unnecessary throughout most of the nadir scene, since both interplanetary and exospheric composition are overwhelmingly dominated by H atoms.  However, OOB emission from non-hydrogen constituents of Earth's atmosphere, along with solar continuum emission reflected from Earth's surface and cloud tops, dominate GCI measurements across Earth's disk and innermost limb.  Masking of the Earth's disk is straightforward and appropriate, since lines-of-sight which intersect the solid Earth, and thus do not transit the full solar-illuminated viewing column, are not suitable for H density retrieval.  However, removal of OOB emission along lines-of-sight that transit the limb near Earth's exobase ($\sim$500 km altitude above the surface) is a prerequisite for H density retrieval in this region, and the algorithms used to estimate these relatively weaker OOB signals are described in Section \ref{sec:oob_algo_head}. Finally, Section \ref{sec:sci_data_processing_l1B_l1C} summarizes the calibration data processing steps which perform the photon background removal and absolute calibration of L1B science images using the estimated photon background signals.   

\section{Payload Description and Concept of Operations}

The primary payload of the Carruthers Geocorona Observatory mission is the GeoCoronal Imager (GCI), which consists of two co-aligned imagers (channels) for simultaneous sensing of exospheric Ly-$\alpha$ \cite{carruthers_lab_cal_paper}. The Narrow Field Imager (NFI), with $1024 \times 1024$  resolution across a $3.6^{\circ}$ Field-of-View (FOV), is designed to provide high spatial resolution (better than 1 arcmin) near the Earth's limb where the exospheric radiance is expected to fall off rapidly. The Wide Field Imager (WFI), with $512 \times 512$ resolution across an $18^{\circ}$ FOV, is designed to capture the entire exosphere with the sensitivity to detect the dimmest Ly-$\alpha$ signals at the outermost extent of the exosphere.  Despite featuring identical UV-intensified CMOS-based detectors, the NFI and WFI are fully independent optical systems with distinct optical paths and wavelength-dependent, open system optical responsivities that achieve wide-band spectral filtering by strongly suppressing far ultraviolet emission below Lyman-$\alpha$ and more weakly suppressing longer wavelength emission into the mid-ultraviolet spectral region. To further refine the open system response and enable out-of-band suppression, estimation, and removal, each channel utilizes an independent but identical 6-position filter wheel, populated as follows:
\begin{itemize}
    \item Open: An unfiltered position providing a transmissivity of unity across all sensitive wavelengths.
    \item LyaN (Narrow): The primary science configuration utilizing an Acton F122-N filter to support narrowband Ly-$\alpha$ transmission (peak transmissivity 22\%).
    \item LyaX (Extra Narrow): A backup Acton F122-XN filter with a lower peak transmissivity of 11\%, designed to mitigate saturation if exospheric Ly-$\alpha$ emissions are brighter than model predictions.
    \item CaF$_2$: A 3mm thick calcium fluoride window acting as a longpass filter. It suppresses Ly-$\alpha$ transmission to isolate longer wavelengths for out-of-band measurement and subtraction.
    \item SrF$_2$: A 3mm thick strontium fluoride window that functions similarly to the CaF$_2$ filter, blocking Ly-$\alpha$ to evaluate out-of-band photon contributions.
    \item Blocked: An opaque aluminum disk providing zero transmissivity at all wavelengths, utilized strictly for instrument calibration purposes.
\end{itemize}

\noindent Out-of-band photon emission characterization and removal is based on sequential, nadir-oriented observations using multiple filters.  Because only the NFI channel features sufficient spatial resolution to capture the large radial gradients in exospheric Lyman-$\alpha$ emission across Earth's limb, OOB image acquisition on nadir and subsequent OOB calibration of exospheric science images is not performed for the WFI channel; inclusion of OOB filters in the WFI filter wheel is intended solely to mitigate the risk of NFI failure on orbit.   

As a standalone spacecraft, the Carruthers Geocorona Observatory has full attitude control for dedicated, off-nadir calibration image acquistion, though the field of regard is limited to 42$^{\circ}$ within the Sun-spacecraft line by both thermal and power positivity constraints \cite{Craig26}.  Off-nadir slewing operations are conducted to support routine flat field and stellar data acquisition for instrument effect correction and optical responsivity calibration, as described by Zhang et al. \cite{Zhang26a} and Zhang et al \cite{Zhang26b}, respectively. Pointing stability around a target boresight direction, estimated to be 13.92 arcsec, is driven mainly by spacecraft jitter and bounds the mechanical smear during an exposure.  Jitter, along with optical diffraction, determine the point spread function (PSF) of the optical system; knowledge of each channel's PSF, whether obtained from lab calibration as used here or from on-orbit observations of stellar point sources, is required for deblurring of limb pixels in support of background OOB emission characterization as well as masking of celestial sources, as described in the next section.  

The baseline concept of operations (ConOps) for GCI commissioning during the extended cruise to L1 after launch involves dedicated slewing to more than 80 off-nadir stellar targets for a comprehensive beginning-of-life sensitivity calibration.  This ConOps plan also features weekly slewing to new stellar targets during the mission's primary science phase as a means of monitoring optical throughput over time.  Non-stellar regions of all off-nadir images, acquired both before and after insertion into halo orbit around L1, exclusively contain IPH emission and are used to constrain the IPH background signal across the full FOV.  Additionally, IPH emission is expected to dominate over the dim exospheric signals beyond $\sim$25 [earth radii] ([$R_E$], where 1 [$R_E$] = 6371 [km]) from Earth. This region is well within the 18$^{\circ}$ FOV of the WFI channel, and the IPH estimation algorithm also incorporates the outer annulus of nadir-oriented WFI images as additional IPH retrieval constraints (see Section \ref{sec:iph_photon_retrieval_algorithm} for details). 
 

The accuracy and precision of the algorithms described in this paper are evaluated using synthetic data produced by numerical, model-based simulations of the expected in-band and out-of-band photon scene and the GCI response to these incident photons.  This numerical simulator supports realistic viewing geometry and image acquisition schemes as well as measurement noise associated with the exospheric target, the background photon scene, and instrument effects.  Relevant details regarding algorithm performance evaluation using synthetic data are provided below as needed; a thorough description of the GCI image simulator is given by Filippini et al. \cite{Filippini26}.    
 
\section{Celestial Photon Background Masking and Interpolation}
\label{sec:celestial_body_alg}

Celestial bodies, specifically stars, the Moon, and the outer planets, introduce non-Ly-$\alpha$ contamination into the telemetered images and must be masked. Because the spatial footprints of most celestial signal sources are small relative to the spatial gradients of IPH and exospheric emission across the FOVs of both channels, the underlying scene signal can be accurately approximated via spatial interpolation.  This section details the algorithms used to identify and mask these celestial sources, as well as the subsequent interpolation methodology used for removing contamination by the smaller, dimmer celestial objects.

\subsection{Stellar signal identification}
\label{sec:stellar_loc_alg}

Precise knowledge of star locations on the image plane is essential for effective masking or interpolation. However, the signal contribution, in [DN], from stars is unpredictable, and compiling a comprehensive catalog of every star warranting removal for each specific channel and filter configuration is impractical. This section outlines an algorithm that identifies stars in each non-blocked image without relying on independent knowledge of stellar locations. By dynamically detecting stars, the algorithm ensures all bright sources are located and allows for a configurable minimum brightness threshold for removal.

The stellar location retrieval algorithm is applied to all L1B images acquired by both imaging channels, following their creation via instrument effect correction of their parent L1A image (see Table \ref{tab:data_product_def_table}). The algorithm first divides the image into a grid of $10$ concentric radial zones and $12$ angular slices, yielding a set of distinct curvilinear regions over which local statistics can be computed. The algorithm then repeats the following steps iteratively:
\begin{enumerate}
    \item Within each curvilinear region, the median pixel value is calculated, excluding any NaN entries imposed by previous rounds of stellar detection processing. These local medians are then projected back onto the image coordinate system to construct a spatial background estimate.
    \item This background estimate is subtracted from the original image to generate a residual image. In this residual frame, stellar sources appear as large positive  outliers against a zero-centered noise floor.
    \item The variance of the residual is calculated within each curvilinear region. These local variance values are subsequently mapped back to the full resolution of the original image grid.
    \item Negative values in the residual image are clamped to zero, and a proxy Signal-to-Noise Ratio (SNR) is derived by dividing the square of each pixel value by the local variance. Pixels exceeding a defined SNR threshold, empirically established at $25$ as described below, are identified as candidate stars. To account for flux distribution due to the instrument Point Spread Function (PSF), these detection points are circularly dilated to encompass the full width of the PSF. The PSF's width was measured prior to launch and analytically broadened to account for on-orbit effects such as spacecraft jitter, thermal distortions, and other pointing errors \cite{carruthers_lab_cal_paper, Filippini26}. The instrument PSF for both channels will be confirmed on-orbit using stellar measurements as point source system inputs.
    \item Identified star pixels are accumulated into a global mask and set to NaN in the data to preclude re-detection. The algorithm iterates until convergence is achieved (defined as no change in the mask between steps) or a maximum of iterations is reached. In the baseline algorithm described here, the maximum number of iterations is initialized to five based on empirically demonstrated optimal performance during pre-flight simulations described below. The efficacy of this user-defined iteration limit will be assessed and potentially updated once operational flight data becomes available post-launch.
\end{enumerate}

Performance evaluation of the stellar signal identification algorithm utilized a numerical image simulator \cite{Filippini26} to generate synthetic L1B images containing stars.  In this case, the input stellar sources were specified using an external database of known stellar photon flux spectra to ensure a wide range of stellar brightnesses for algorithm testing; however, as noted above, identification of stellar contamination does not depend on prior knowledge of stellar location or their spectral characteristics.  Stars were simulated against a realistic, macroscopic background consisting solely of IPH and exospheric emission, and the simulator employs the same analytically broadened PSF described in step 4 of the algorithm, which explicitly incorporates the convolution of sensor and vehicle effects.  

Algorithm  performance was assessed in terms of excess signal present after masking, relative to the known ground-truth modeled scene in the absence of stars.  This evaluation established initial, pre-launch values for the user-defined algorithm settings, such as the discretization of the image into curvilinear regions, the threshold SNR for stellar detection in step 4, and the maximum number of iterations needed to achieve sufficient detection of dim stars in step 5. The chosen parameter settings successfully removed stellar contamination in off-nadir and outer exospheric regions dominated by relatively flat and dim Lyman-$\alpha$ emission but was less successful for dim stars near Earth's limb, where the exospheric signal variance is larger owing to strong radial gradients across the curvilinear region used to define stellar-background signal statistics. Once on-orbit, the user-defined parameters which govern the masking algorithm will be evaluated and adjusted as needed, based on observed stellar fields and nadir scene gradients.

\begin{figure}[htbp]
    \centering
    \includegraphics[width=\textwidth]{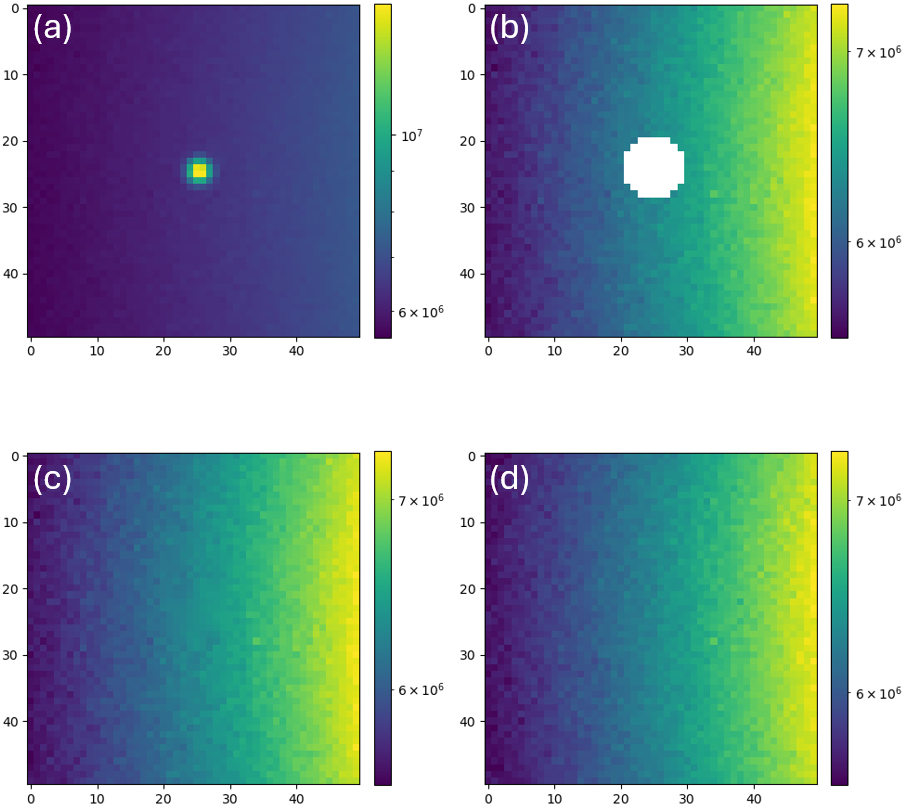}
    \caption{Example stellar signal identification, masking, and interpolation.  Panel (a): Noisy realization of a synthetic NFI L1B image, clipped around a representative modeled star.  Panel (b): The same synthetic data shown in panel (a), but with pixels containing identified stellar photon contamination masked.  Panel (c): The same synthetic data shown in panel (a), but with the signal within masked pixels replaced by values interpolated from the surrounding image. Panel (d): A baseline simulation of the same region generated entirely without the target star. In all panels, the colorbar depicts signal rate in [DN/s].}
    \label{fig:star_mask}
\end{figure}

Figure~\ref{fig:star_mask} displays an example of the stellar identification and masking algorithm performance, which demonstrates the success of the approach and the spatial extent of a typical stellar mask applied to the input scene.  The size of voids associated with stellar sources varies with both the inherent stellar brightness as well as the channel/filter configuration used to measure it.  As a result, the star masking algorithm is applied independently to each L1B image before further processing.  

\subsection{Moon and Planet Masking}
\label{sec:moon_and_planet_loc}

Effective masking and interpolation of the Moon and outer planets require precise localization on the image plane. Unlike stars, which require an on-the-fly detection algorithm due to their sheer number, major solar system bodies can be tracked and masked using direct calculation. Their cataloged ephemerides \cite{astropy2022astropy} allow their coordinates and apparent angular sizes to be deterministically predicted for each channel, even accounting for motion blur during the 30-60 minute integration times of telemetered images.

After instrument effect correction, all L1B image pixels contaminated by signals from the Moon or outer planets will be clipped and replaced with NaNs in further image processing. As with stellar signal masking described above, the clipping dimensions are conservatively defined to fully accommodate expected spatial blurring in terms of the known instrument PSF, which will be assessed using on orbit data and updated as needed.  

\subsection{Star/Moon/Planet Interpolation}
\label{sec:hole_interp}

Following the identification and masking of pixels containing signal from celestial objects, the resulting images contain numerous data gaps, particularly when viewing the relatively dense star fields associated with the biannual transits of the galactic equator across the nadir field-of-view. Rather than propagating these voids through subsequent processing stages, it is preferable to populate them via interpolation using adjacent pixel data. The approach is well supported by the generally smooth nature of the observed scene, which ensures high interpolation accuracy as long as the interpolated region is sufficiently small.  

The reconstruction utilizes the Radial Basis Function (RBF) interpolator provided by the scipy.interpolate package \cite{virtanen2020scipy}. The interpolation is constrained to the nearest $100$ neighboring pixels and incorporates a third-degree polynomial into the basis to mitigate overfitting to noise.  Data voids with a total area exceeding an empirically derived threshold are preserved, while smaller gaps undergo interpolation. The baseline interpolation size threshold was determined through analysis of an ensemble of simulated L1B images produced using the numerical instrument and scene models described above \cite{Filippini26}. Ten synthetic masked holes were placed randomly within the FOV of a given noisy realization of L1B data, with hole sizes varying from 80 to 540 pixels.  The interpolated signals across the voided regions were evaluated in terms of their statistical agreement with the unmasked ``ground truth" signals across the same region.  The baseline root mean squared error (RMSE) among the different noisy NFI and WFI realizations is 5.6\% and 2.2\%, respectively, which corresponds to the variation associated with measurement noise over a given masked area independent of potential interpolation error.  Interpolation-induced error (also in terms of RMSE) is statistically identical to this baseline value for NFI (WFI) holes smaller than 150 (80) pixels and increases from 5.6\% to 6.9\% (from 2.2\% to 3.0\%) for the maximum hole size considered.  The mask interpolation algorithm implemented in the Carruthers data processing pipeline adopts a maximum void size threshold of 250 pixels for both channels, corresponding to a total RMSE of 5.9\% and 2.5\%, respectively.  This threshold is sufficient to effectively remove all dim star and outer planet signals from L1B images, while preserving image continuity for scientific analysis. As a user-defined parameter, the efficacy of this baseline threshold will be evaluated in terms of its on orbit performance and updated as deemed necessary.  

An example of baseline interpolation over a masked star is shown in Panel (c) of Figure~\ref{fig:star_mask}, using the same test case synthetic data shown in panels (a) and (b).  The masked pixels are repopulated with values consistent with the surrounding background, effectively excising the stellar source from the image. Note that the restored area appears marginally smoother than the corresponding region in the starless reference image (shown in panel (d)), as the deterministic interpolation inherently cannot replicate the stochastic noise profile of the surrounding image data.

\section{IPH Photon Retrieval}
\label{sec:iph_photon_retrieval_algorithm}

\begin{figure}[htbp]
    \centering
    \includegraphics[width=0.8\linewidth]{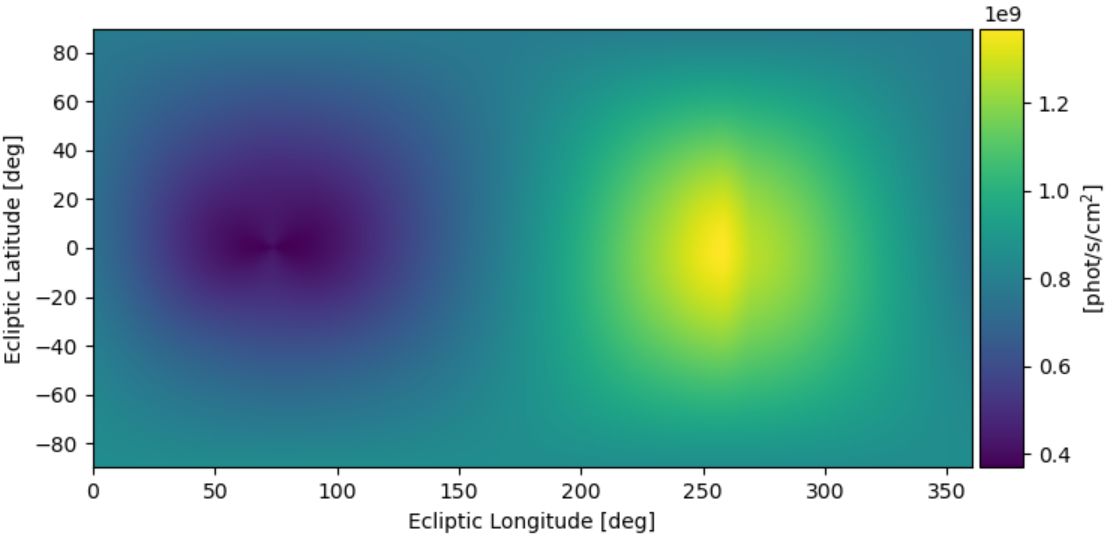}
    \caption{Example IPH Ly-$\alpha$ radiance distribution derived using the Pryor model \cite{IPHModelPryor2013} under solar maximum conditions (specifically assuming a solar flux index of $F_{10.7} = 170$ [sfu]).  The modeled distribution is depicted as a function of latitude and longitude in Heliocentric Mean Ecliptic coordinates, a spherical coordinate system with the mean ecliptic plane of J2000.0 as the reference plane. Longitude is defined as the angular coordinate in the ecliptic plane, measured from the vernal equinox, while latitude is defined as the angular distance of the object above or below the ecliptic plane. Radial distance can be optionally specified, but only the directional components are relevant for IPH mapping.}
    \label{fig:iph_map}
\end{figure}

The in-band photon background is dominated by resonant scattering of solar Ly-$\alpha$ photons by InterPlanetary Hydrogen (IPH) atoms of interstellar origin. These neutral H atoms penetrate and flow through the solar system, though their velocity distribution is altered by charge exchange at the heliospheric interface with the interstellar medium.  Variations in scattering H atom velocity affect scattered emission radiance via Doppler shifting away from resonance.  Consequently, the global IPH scene exhibits a characteristic pattern associated with the heliosphere's motion through interstellar space \cite{lallement1993deceleration}.  Figure \ref{fig:iph_map} depicts a physics-based model simulation of the IPH radiance distribution corresponding to solar maximum conditions in heliocentric coordinates.   

While the global IPH radiance pattern is time invariant over the relatively short lifetime of the Carruthers mission, the measured IPH radiance along a given line-of-sight (LOS) look direction varies linearly with the incident solar Lyman-$\alpha$ flux that excites the emission at the time of the observation.   Temporal variation of solar photon flux is significant over both the 11-year solar cycle and the $\sim$27-day solar rotation period, where the variation is associated with the spatially localized, transient evolution of sunspots and solar flares.  Our approach to IPH estimation and subtraction along exospheric crossing lines-of-sight takes advantage of the invariant nature of the IPH radiance pattern while also accounting for its localized temporal variation in response to changing solar Lyman-$\alpha$ flux. 

The basic steps of IPH background estimation and subtraction are summarized as follows; a more complete description of each step is provided in the subsections below.  The first step is the construction of a reference map of the global IPH radiance distribution, based on normalized aggregration of all GCI observations of isolated IPH emission.  This reference map represents the time-invariant spatial dependence of IPH radiance, and the normalization is chosen to correspond to the IPH signal rate (in [DN/s]) as observed by the WFI LyaN configuration, which dominates the data volume. In order to remove the short-term variability associated with solar illumination, every GCI image used to construct the reference map is normalized by the solar Ly-$\alpha$ flux during IPH image acquisition, assumed known as described below, as well as the optical responsivity of the channel and filter combination used to acquire the IPH image, relative to the WFI LyaN configuration.  The next step is to use this reference map to estimate the IPH signal along exospheric-crossing lines-of-sight in a target L1B science image (i.e., nadir viewing with a narrowband Lyman-$\alpha$ filter).  The reference map signal value in the pixel of interest is scaled by the target instrument's responsivity relative to the WFI LyaN configuration and by the specific solar Lyman-$\alpha$ flux at the time of science image acquisition.  Lastly, the estimated background IPH signal is subtracted from the measured L1B signal in that pixel, yielding isolated exospheric signal suitable for absolute intensity calibration into physical radiance units as the last step of the L1B $\rightarrow$ L1C data processing pipeline.  The efficacy of this algorithm relies on knowledge of two key quantites: (1) the incident solar Ly-$\alpha$ flux at the time of image acquisition, discussed further in Section~\ref{sec:iph_map_retrieval_algorithm} below, and (2) the relative optical responsivity at Ly-$\alpha$ between all channels/filters and WFI/LyaN, discussed next. 

\subsection{Relative Responsivity Retrieval at Ly-$\alpha$}
\label{sec:cross_channel_responsivity}

Knowledge of absolute calibration of all channel/filter settings based on observation of calibration stars was discussed in Zhang et al. \cite{Zhang26c} and can be used to obtain the relative Ly-$\alpha$ responsivity required for IPH photon background correction. However, relative photometry can be performed at much higher precision and accuracy by leveraging the fact that vast regions of the scene in all GCI images acquired on orbit contain only Ly-$\alpha$ photons, whether it be produced by Earth's exosphere or by the IPH.  This latter approach, described in this section, is used for IPH estimation and subtraction for the Carruthers data processing pipeline.

The Ly-$\alpha$ responsivity of a given channel/filter configuration, relative to the WFI LyaN imaging mode, is retrieved using instrument-effect corrected L1B images, excluding regions where non-Lyman-$\alpha$, out-of-band contamination is present.  Specifically, pixels affected by stars, the Moon, or any outer planets are masked, as well as those within a user-defined distance from nadir, chosen conservatively here as $2R_E$ to account for PSF-associated broadening of unknown out-of-band emissions from the disk and lower atmosphere into pixels viewing the upper limb.  The NFI FOV encompasses LOS tangent point radial distances out to $\sim7.4R_E$ (the WFI FOV extends to $\sim35R_E$); therefore, excluding the pixels within $2R_E$ removes only $\sim 8\%$ of the total NFI observational area (and far less for the WFI), preserving the vast majority of the science data as cross-calibration constraints.

The relative Ly-$\alpha$ responsivity retrieval algorithm is designed as a generalized cross-calibration routine. It derives a relative scaling factor by computing the ratio of average signal rates [DN/s] among non-masked pixels in the FOV between any two distinct channel/filter configurations observing an identical target scene. To maintain operational flexibility, a valid configuration pairing can consist of either: (1) simultaneous acquisitions across different channels (e.g., comparing WFI to NFI), or (2) sequential acquisitions from a single channel utilizing different filters (e.g., comparing NFI/Open to NFI/LyaN), which necessitates an operational assumption of temporal scene stability.  The different sizes of the FOV must be considered when recovering relative responsivity across different channels; in this case, only WFI pixels sampling regions of the scene that are also sampled simultaneously by the NFI are used in the calculation of signal rate, in [DN/s], for the WFI image. 

The algorithm was validated using the numerical image simulator \cite{Filippini26} to generate a set of $500$ synthetic, instrument-effect-corrected L1B images, corresponding to approximately one week of baseline on-orbit operations. To assess the algorithm against realistic on-orbit conditions, these simulated scenes featured modeled IPH and geocoronal Ly-$\alpha$ emission subject to standard stochastic sensor effects, including realistic noise sources. Absolute responsivity values derived from ground-based lab calibrations \cite{carruthers_lab_cal_paper} were utilized to convert these physical backgrounds into the simulated instrument response. Images were generated for each channel using the Open and LyaN filters with an integration time of $30$ minutes, matching the baseline on-orbit image acquisition scheme. After masking out-of-band contaminated pixels, all six possible cross-channel and cross-filter permutations (e.g., WFI Open vs. WFI LyaN, WFI Open vs. NFI Open, etc.) were processed through the retrieval algorithm. The retrieved relative Ly-$\alpha$ responsivity was then compared against the relative Ly-$\alpha$ responsivity obtained from ground calibrations in terms of residual error. The algorithm performs well under these noisy conditions, yielding a negligible mean error and a low standard deviation of $0.25\%$. This high precision, particularly when compared to stellar-based absolute responsivity methods, is physically driven by the sheer volume of the data: the relative responsivity is a ratio of two averages, each derived from $\sim700,000$ independent, statistically distributed pixels in the NFI (and $\sim10,000$ in the WFI). Consequently, the signal-to-noise ratio of the macroscopic average is extraordinarily high, enabling highly accurate cross-calibration despite pixel-level variance.

\subsection{Normalized IPH Reference Map Construction}
\label{sec:iph_map_retrieval_algorithm}

Carruthers routinely acquires wide-field coverage of the IPH scene isolated from exospheric Lyman-$\alpha$ signal contributions, namely in non-masked regions of all off-nadir NFI and WFI images and in the outer annuli of all WFI science images beyond $25R_E$ from nadir.  This outer radial boundary is a user defined input parameter, whose baseline value of $25R_E$ is chosen here based on modeled expectations.  Specifically, the outer exospheric ground-truth H density model (see Figure 3 in \cite{Filippini26}) produces optically thin Ly-$\alpha$ emission radiance that is $\sim 4.18 \times 10^6 \pm 0.55 \times 10^6$ [photon/s/cm$^2$] at a tangent point radial distance of $25R_E$, which is $\sim 1\%$ of the minimum IPH radiance of $\sim 4 \times 10^8$ [photon/s/cm$^2$] in the modeled IPH scene.   

As with the relative responsivity algorithm described above, the normalized IPH retrieval algorithm ingests L1B images from both channels, restricted to the open, LyaN, and LyaX filters (the OOB filters have negligible Ly-$\alpha$ throughput). Prior to ingestion, celestial sources such as the stars, the Moon, and the outer planets are either masked or interpolated using the algorithm described in Section \ref{sec:hole_interp}.  These OOB-filtered L1B data are used to populate an IPH reference map in Heliocentric Mean Ecliptic latitude and longitude at a $0.2^\circ \times 0.2^\circ$ resolution. A corresponding IPH total measurement time map is constructed alongside the IPH reference map, which stores the total measurement time each pixel of the IPH map has received, in units of [sec], with the same resolution in the same coordinate system.

The algorithm begins by normalizing input images by the time-dependent solar Lyman-$\alpha$ flux at the center of the emission line, in units of [photons/s/cm$^2$/{\AA}] (see below), and by the cross-calibration factor of the input image relative to the WFI/LyaN responsivity (see the previous section for details of GCI cross calibration retrieval on orbit).  Normalization by the WFI/LyaN cross calibration factor ensures that binned signal rates in the IPH reference map correspond to those measured using the WFI/LyaN configuration.  The plate scale of the image is calculated based on the known camera viewing geometry, and relevant pixels are then extracted and spatially downsampled to the IPH map resolution via linear interpolation.  Data integration follows a time-weighted averaging scheme: if a map pixel is unpopulated, the new measurement is weighted by its integration time and stored. Conversely, if a pixel contains existing data, the algorithm retrieves the stored value, reconstructs the total integrated signal, adds the new time-weighted measurement, renormalizes by the updated total integration time, and updates the corresponding total measurement time map. This process yields a cumulatively averaged, normalized IPH signal rate for each pixel in units of [(DN/s) / ($10^{11}$ phot/s/cm$^2$/{\AA}) ], with the $10^{11}$ factor removed to ensure optimal precision retention and computational convenience for downstream users.   

Knowledge of the incident solar Lyman-$\alpha$ line center flux at the time of IPH data acquisition is required for ensuring that short-term temporal variability in GCI-observed IPH radiance does not influence the spatial structure of the IPH scene specified in the reference map.  We retrieve this knowledge on orbit as a by-product of exospheric H density retrieval using NFI observations of Earth's upper limb, an approach which exploits the linear dependence of the resonantly scattered emission radiance on the incident solar flux.  Owing to this linearity, exospheric H density local to the exobase, $\sim$500 km altitude, is constrained solely by the observed scale height of the limb radiance profile (i.e., its  dependence on radial distance above the limb) and not by its radiance in absolute terms \cite{bishop2001thermospheric, waldrop_lyman_a_airglow_emission, joshi2018ionospheric, Joshi26a}.  After retrieval of the inner exospheric H density in the Carruthers L1C$\rightarrow$L2 pipeline, a second optimization routine is performed to retrieve the line-center solar flux as that required to reconcile the measured absolute radiances at the limb, subject to absolute responsivity calibration, with those modeled in terms of the independently retrieved H density (see \cite{Joshi26a} for further details).  

A recursive dependency between H density retrieval and IPH background calibration exists because IPH photon retrieval and removal is itself a prerequisite for calibrating the radiance data used as input to the H density retrieval. To resolve this recursiveness during on-orbit operations, the IPH mapping algorithm utilizes solar flux estimates from the most recent data processing cycle, which introduces a typical latency of 24 hours. However, given that solar Ly-$\alpha$ flux exhibits stability over timescales of approximately 14 days (half a solar rotation) \cite{woods2000solarlyafluxvar}, this latency is considered negligible, particularly since each binned region of the IPH reference map is constrained by multiple days of GCI observations.  Once on orbit, we will assess the reliability of GCI-based estimates of solar Lyman-$\alpha$ flux using independent, line-integrated solar flux data acquired from earth-orbiting spectrographs \cite{machol2019improved} and, when solar viewing is favorable, from the Lyman-$\alpha$ photodiode that comprises the mission's secondary science payload, the Carruthers Student Solar Monitor (COSSMo) \cite{Naldoza2025COSSMo}.  

Performance assessment of the IPH reference map construction algorithm relied on a manually constructed, semi-realistic observation schedule designed to emulate mission conditions. The timeline for synthetic data production encompasses the science commissioning phase (November 7th, 2025 to January 1st, 2026), featuring dedicated off-nadir slews to $80$ distinct stellar targets for initial responsivity calibration, followed by nominal science operations extending through November 7th, 2026. The post-commissioning schedule incorporates $11$ flat-field slews and $70$ new stellar target acquisitions distributed throughout the period after January 1st, 2026, with the remaining intervals allocated to nadir observations, though produced at daily cadence of hour-long integration instead of at the hourly cadence of WFI images on orbit.   All synthetic instrument-effect-corrected images (L1B equivalent) correspond to beginning-of-life conditions and were generated using the numerical image simulator described in Filippini et al. \cite{Filippini26}, incorporating realistically retrieved solar Ly-$\alpha$ flux from noisy, synthetic NFI limb data. Celestial sources such as the stars, the Moon, and the outer planets were either masked or interpolated using the algorithm described in Section \ref{sec:hole_interp} before ingestion into the IPH mapping algorithm.

\begin{figure}[htbp]
    \centering
    \includegraphics[width=0.95\linewidth]{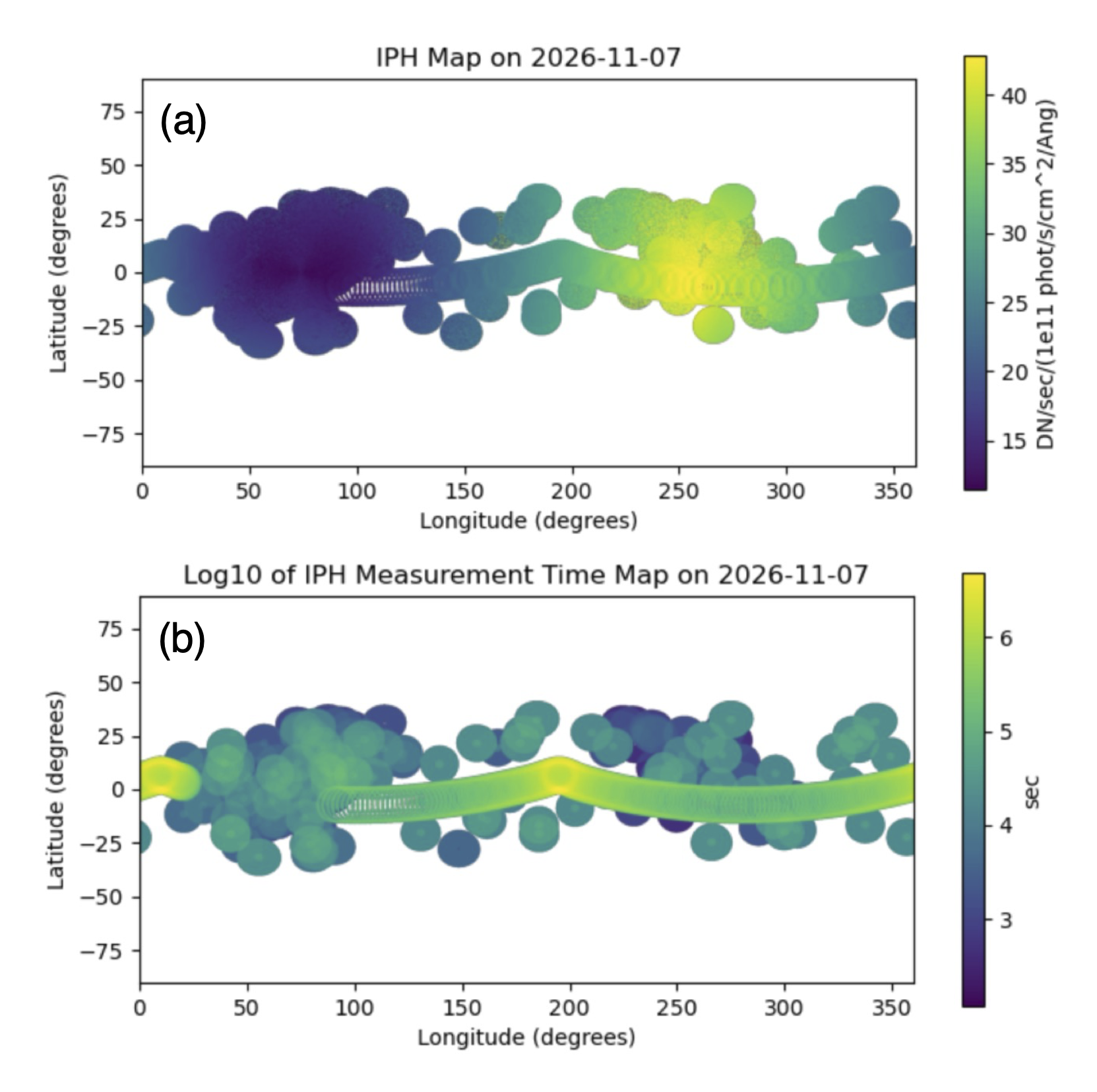}
    \caption{IPH background signal retrieval after one year of semi-realistic science operations. Panel (a) depicts the normalized IPH signal rates corresponding to the WFI/LyaN configuration, while panel (b) depicts the total measurement time, on a log10 scale, of the data used to constrain each $0.2^{\circ}\times 0.2^{\circ}$ map element.  The first 55 days are spent collecting stellar observations, hence the density of off-nadir pointing on the leftmost side of the plot. The overlapping rings near the center of the image correspond to WFI nadir image annuli, though their discrete nature is an artifact of the reduced temporal resolution of the synthetic data set. White regions indicate no measurement of IPH at those pixels.}
    \label{fig:iph_map_example}
\end{figure}

Figure \ref{fig:iph_map_example}(a) presents the reconstructed IPH map derived from this year of synthetic data, while Figure \ref{fig:iph_map_example}(b) depicts the spatial distribution of accumulated measurement time associated with map data constraints. Relative to the smooth distribution in modeled IPH radiance, the map values exhibit spatial variation that is correlated with observation type: regions constrained only by off-nadir stellar images show higher variance due to the short integration times characteristic of these observations. In contrast, the overlapping rings in the center, derived from WFI annuli around Earth nadir, represent the smoothest portions of the dataset, reflecting the large amount of time allocated to nadir pointing. The discrete annular structures visible in the IPH map and total measurement time map are artifacts of downsampling the spacecraft ephemeris and viewing geometry to a daily resolution to reduce computational load; operational on-orbit maps will show a spatially smoother distribution.

\begin{figure}[htbp]
    \centering
    \includegraphics[width=0.65\linewidth]{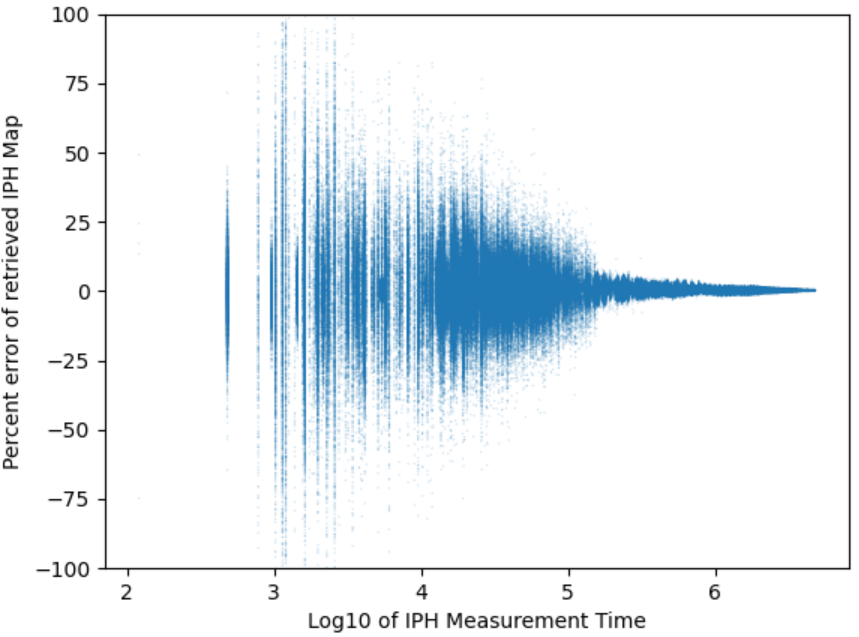}
    \caption{Scatter plot of percent error on retrieved IPH map vs. log10 IPH total measurement time map across all pixels of the IPH map after one year of realistic science operations. As expected, the percent error converges to 0\% as measurement time increases.}
    \label{fig:iph_perc_err_scatter}
\end{figure}

The relationship between the integration time of the data constraints used to reconstruct the IPH map and fidelity to the ground-truth IPH scene model is more clearly illustrated in Figure \ref{fig:iph_perc_err_scatter}.  The figure depicts the residual percent error between the retrieved and ground-truth IPH map values as a function of total measurement time.  Significant deviations from ground-truth are strictly confined to regions with low total accumulated signal, driven operationally by short localized integration times, which is entirely consistent with statistical expectations. The statistical error funnel tightens to within $\pm 5\%$ error after $\sim 10^6$ seconds of accumulated measurement time, a threshold which is routinely achieved for exospheric-crossing lines of sight over the full course of the mission under the baseline concept of operations featuring hourly WFI image acquisition. Furthermore, while isolated spatial bins with lower accumulated integration times exhibit higher native variance, the raw IPH reference map is subsequently processed through a spatial polynomial interpolator (detailed in the next section) prior to operational use, effectively neutralizing localized statistical noise.  

\subsection{IPH Estimation Within Science Images}
\label{sec:iph_retrieval_algorithm}

The background signal contribution of IPH photons, in units of [DN], to be subtracted from nadir science images in the L1B$\rightarrow$L1C pipeline, is determined through systematic scaling and fitting of the normalized IPH reference map constructed as described in the previous section. First, the IPH map is renormalized by the most recent available solar Ly-$\alpha$ flux measurement and by the WFI LyaN cross-calibration factor. The map is then linearly interpolated to match the target image coordinates. Finally, to mitigate noise and to bridge coverage gaps in under-sampled regions, a weighted 2D polynomial fit is applied to derive the final IPH background signal rate in [DN/s]. This fit utilizes an order $3$ polynomial for WFI images and order $2$ for NFI images, reflecting their distinct FOVs. The fitting weights are derived from the IPH total measurement time map, ensuring that pixels representing longer integration time, and thus higher statistical reliability, are prioritized.

The IPH interpolation algorithm performance was evaluated by using the numerical image simulator to generate a raw (L0) LyaN image on each channel corresponding to a realistic, pre-flight beginning-of-life case study. The LyaN image had an integration time of $30$ minutes for NFI and $60$ minutes for WFI, designed to mimic the nominal parameters of planned nadir-pointed science observations. These simulated L0 images were processed through all instrument effect corrections described in Zhang et al. \cite{Zhang26a}, with supporting images generated using the numerical image simulator as necessary. Finally, the stars, the Moon, and the outer planets are masked out or interpolated. The resulting calibrated images were then processed by the background IPH signal  retrieval algorithm using the reference IPH map described above. The mean percent error on the retrieved IPH signal was found to be $-0.13\%$ and $0.4\%$ for the NFI and WFI channels respectively, while the standard deviation is found to be below $1.4\%$ for both channels. The science requirements traceability matrix mandates an IPH background estimation accuracy of $30\%$ to achieve baseline mission success \cite{Waldrop26}. Because the evaluated mean bias and standard deviation both fall well below this threshold, the proposed IPH map construction and removal algorithms successfully satisfy this operational requirement.

\section{Out-Of-Band Photon Retrieval}
\label{sec:oob_algo_head}

The Out-of-Band (OOB) photon background constitutes the dominant source of background signal  contribution in [DN] in pixels proximal to Earth's disk and limb below the exobase ($\sim400-500$ km altitude). Accurate subtraction of the OOB background from NFI nadir images is a critical prerequisite for using measured Lyman-$\alpha$ radiance along exobase-crossing lines-of-sight to constrain the retrieval of thermal H density based on radiative transfer (RT) model inversion \cite{Joshi26a}.  We note that masking of OOB-affected pixels is straightforward, and inner exospheric H density instead could be retrieved without direct exobase measurement constraints.  However, the inherent RT model assumption of H thermalization is increasingly unphysical with increasing distance from the neutral atmospheric populations (mainly O and N$_2$) responsible for imposing the collisional thermalization \cite{Waldrop26}.  Thus, the motivation to pursue OOB estimation and subtraction is to avoid potential thermal H retrieval bias associated with the effects of Lyman-$\alpha$ scattering from a non-thermal H population at high altitudes.        

Because OOB emission exhibits large spatial gradients near the limb, its accurate estimation requires effective spatial deblurring prior to signal extraction. Consequently, this section first details the required spatial deblurring algorithm before outlining the complete OOB photon background retrieval methodology.   In the description that follows, upper-case letters (English or Greek) denote random variables, upper-case letters with an arrow, such as $\vec{X}$, denote random vectors, bold upper-case letters (English or Greek) denote matrices or sets, lower-case letters with an arrow, such as $\vec{x}$, denote vectors, and lower-case letters denote constants or indices. Variables $\epsilon$ that are a function of wavelength are denoted $\epsilon(\lambda)$, while variables $\epsilon$ that are a function of pixel position are denoted $\epsilon(i, j)$, where $i$ denotes the row index ($y$ direction) and $j$ denotes the column index ($x$ direction).

\subsection{Spatial Deblurring}
\label{sec:deblur_algorithm}

Incoming photon flux is convolved with the instrument Point Spread Function (PSF), $h(i,j)$. Previous analyses have focused on sources where PSF mitigation is straightforward: point or small circular sources (stars, the Moon, and outer planets) are managed by extending radial clipping, while diffuse sources with slow spatial variation (such as IPH) require no specific PSF correction. In contrast, the OOB photon background exhibits large spatial gradients driven by the abrupt termination of Earth's disk at $1$ Earth-radii (Re) and the rapid exponential decay of upper atmospheric airglow radiances. Accurate OOB retrieval therefore requires effective image deblurring. However, traditional methods, such as Wiener deconvolution, proved insufficient for reconstructing these sharp spatial gradients. Consequently, a specialized deblurring algorithm was developed to address these limitations. This section details the advanced method, which yields substantial improvements in OOB retrieval accuracy as demonstrated later in this section.

The deblurring algorithm uses a variation of the plug-and-play framework described in Venkatakrishnan et al. \cite{venkatakrishnan_plug_and_play}. Unlike classical deconvolution, the plug-and-play framework uses denoising algorithms as a prior for the final image rather than applying standard regularization methods like Tikhonov regularization. A wavelet-based thresholding technique is used for denoising in the deblurring algorithm presented in this section. Let $\vec{y}$ represent the image to be deblurred and $\boldsymbol{H}$ represent the linear operator that convolves input $\vec{x}$ with the PSF $h(i,j)$. The deblurring algorithm initializes $\vec{x}_0$ and $\vec{z}_0$ to zero, then subsequently executes approximately 40 iterations of the following update sequence:
\begin{enumerate}
    \item $\vec{x}_k = \underset{\vec{x}}{\argmin} \| \vec{y}-\boldsymbol{H}\vec{x} \|_2^2 + \gamma_1\|\vec{x}-\vec{z}_{k-1} \|_2^2$
    \item $\vec{z}_k = \text{WaveletDenoiser}(\vec{x}_k, \frac{\gamma_2}{\gamma_1})$
\end{enumerate}

In this formulation, $\gamma_1$ and $\gamma_2$ serve as tunable hyperparameters governing algorithmic convergence, and the final deblurred output is stored as $\vec{x}_{40}$. The minimization step for $\vec{x}_k$ is solved numerically via gradient descent utilizing the ADAM optimizer \cite{kingma2017adam}. To maximize computational efficiency through parallelization, the algorithm is implemented in PyTorch \cite{paszke2019pytorch}.

As the primary objective of the deblurring algorithm is to minimize error in the OOB photon background retrieval, the algorithm is not validated in isolation. Instead, its performance is assessed as an integrated component of the OOB retrieval pipeline in the next section.

\subsection{Out-of-band Photon Retrieval}

Throughout this section, let $r_{ij}$ denote the Euclidean distance between the center of the Earth and the tangent point of the line of sight (LOS) of each pixel $i,j$ to the surface of the Earth, in units of Earth-radii (Re). The tangent point is defined as the point of closest approach to the spherical body. Let $\boldsymbol{D}$ be the set of pixels that satisfy $r_{ij} < 1$, which implies that the pixels in $\boldsymbol{D}$ span the Earth itself. The area of the image covered by the pixels in $\boldsymbol{D}$ is referred to as Earth's disk. Let $p_{ij}, q_{ij}$ be the pixel indices of the closest pixel in $\boldsymbol{D}$ to pixel $i, j$ (with respect to Euclidean distance on the image plane).

To facilitate OOB photon background estimation and removal, the NFI channel executes a sequential observation consisting of 30-minute exposures using the LyaN, Open, CaF$_2$, SrF$_2$, and LyaN filter, all pointed at Earth nadir. Because the instrument utilizes a single optical path with a sequential filter wheel, simultaneous multi-filter observation is impossible. Consequently, decoupling the specific contributions from Ly-$\alpha$, O1304, O1356, and other wavelengths via known filter transmissivities requires the operational assumption of temporal stability throughout the 2.5-hour sequence. While thermospheric and auroral sources can exhibit natural variability on these timescales \cite{eastes2019global}, treating the scene as static is a necessary hardware-driven constraint for this retrieval algorithm. The two LyaN images are stacked prior to processing to enhance the Signal-to-Noise Ratio (SNR). 

Because thermospheric and auroral emissions at 1304 Å and 1356 Å can exhibit temporal variability over the 2.5-hour sequence, the bracketing LyaN exposures serve a critical diagnostic function. By quantifying the temporal drift between the initial and final LyaN frames, the pipeline can bound the error introduced by the assumption of temporal stability, ensuring data integrity during geomagnetically dynamic periods.

For exobase Hydrogen density retrieval, OOB characterization is strictly required within the altitude range of 500–1500 km ($1.078 R_E < r_{ij} < 1.4 R_E$); this specific set of pixels is denoted as $\boldsymbol{J}$. Because the mission concept of operations excludes the use of WFI data at these altitudes, the OOB signal in WFI images is not considered in this analysis.

\begin{table}[htbp]
    \centering
    \renewcommand{\arraystretch}{1.3}
    \begin{tabularx}{\textwidth}{llX}
        \toprule
        \textbf{Symbol} & \textbf{Units (if applicable)} & \textbf{Definition} \\
        \midrule
        $b$ & & Filter \\
        $\boldsymbol{D}$ & & Set of pixels satisfying $r_{p_{ij}q_{ij}} < 1$ \\
        $\text{deblur}(\cdot)$ & & Algorithmic deblurring operator described in Section \ref{sec:deblur_algorithm}. \\
        $h(i, j)$ & & 2D point spread function (PSF) blurring kernel \\
        $i, j$ & & Pixel indices in the Field of View (FOV) \\
        $i_m(i, j, t)$ & [photons/s/cm$^2$] & Intensity of discrete line emission $m$ \\
        $\boldsymbol{J}$ & & Set of pixels where OOB characterization is required ($1.078 < r_{ij} < 1.4$) \\
        $\ell(\lambda, i, j, t)$ & [photons/s/cm$^2$/{\AA}] & Scene spectral radiance. Subscript `exo' indicates exospheric spectral radiance, subscript `OOB' indicates out-of-band spectral radiance, and subscript `bkgd' indicates background spectral radiance. \\
        $n_{\text{frame}}$ & [frames] & Total number of image frames \\
        $p_{ij}, q_{ij}$ & & Pixel indices of the closest pixel in $\boldsymbol{D}$ mapped from $i, j$ \\
        $r_b(\lambda)$ & & Absolute responsivity of filter $b$ at wavelength $\lambda$ \\
        $r_{ij}$ & [Re] & Radial distance corresponding to pixel $i, j$ \\
        $S_{\text{FOV}, b}$ & [DN] & Signal in the FOV for OOB input through filter $b$ \\
        $t$ & [s] & Time \\
        $t_k$ & [s] & Start time of frame $k$ \\
        $t_{\text{frame}}$ & [s/frame] & Duration of a single image frame \\
        $t_{\text{int}, b}$ & [s] & Integration time of the image taken with filter $b$ \\
        $\delta(\lambda - \lambda_m)$ & [{\AA}$^{-1}$] & Dirac delta distribution centered at $\lambda_m$ \\
        $\lambda$ & [\AA] & Wavelength \\
        $\Omega$ & [sr] & Pixel solid angle \\
        \bottomrule
    \end{tabularx}
    \caption{Nomenclature and parameters for the OOB retrieval and removal algorithm.}
    \label{tab:oob_symbols}
\end{table}

Table \ref{tab:oob_symbols} contains a list of symbols and definition used in the remainder of this section. The Out-of-Band (OOB) retrieval algorithm processes $30$-minute exposures of a static scene on the NFI channel, utilizing the filter sequence and image stacking protocol described previously. Input images require extensive preprocessing: instrument effects must be corrected (see Zhang et al. \cite{Zhang26a} for details), IPH photon backgrounds must be removed (see Section \ref{sec:iph_photon_retrieval_algorithm}), stellar sources must be masked and interpolated over (see Section \ref{sec:hole_interp}), and the Moon and outer planets must be masked (see Section \ref{sec:moon_and_planet_loc}). Furthermore, the algorithm relies on knowledge of absolute responsivities, $r_b(\lambda)$ for filter $b$, for all filters at the $1216${\AA}, $1304${\AA}, and $1356${\AA} emission lines (the absolute responsivity retrieval algorithm is detailed in Zhang et al. \cite{Zhang26c}).

The mean of each pixel in the FOV for non-blocked images was derived Filippini et al. \cite{Filippini26}. Removing instrument effects, plugging in the definition of $r_b(\lambda)$ from Zhang et al. \cite{Zhang26c}, and expanding yields the following expression:
\begin{equation}
    \mathbb{E}\left[S_{\text{FOV}, b}(i, j)\right] = \sum_{k = 1}^{n_{\text{frame}}} \frac{\Omega}{4\pi} \int_{t_k}^{t_k + t_{\text{frame}}} \int r_b(\lambda)(\ell(\lambda, i, j, t) * h(i,j)) d\lambda dt
\end{equation}
Next, the scene model described in Filippini et al. \cite{Filippini26} can be substituted for $\ell(\lambda, i, j, t)$. Since IPH, stars, the Moon, and the outer planets have already been removed, the only two terms remaining in $\ell(\lambda, i, j, t)$ are $\ell_{\text{exo}}(\lambda, i, j, t)$ and $\ell_{\text{OOB}}(\lambda, i, j, t)$. For analytical convenience, the constituent terms are consolidated and subsequently re-partitioned into two new terms. The first term isolates the discrete line emissions (H1216, O1304, and O1356), characterized by Dirac delta distributions in wavelength. This component is explicitly expanded as a summation indexed by $m$. The second term, representing the remaining non-discrete sources, is aggregated into the background spectral radiance $\ell_{\text{bkgd}}(\lambda, i, j, t)$.
\begin{multline}
    \mathbb{E}\left[S_{\text{FOV}, b}(i, j)\right] = \sum_{k = 1}^{n_{\text{frame}}} \frac{\Omega}{4\pi}\\ \int_{t_k}^{t_k + t_{\text{frame}}} \int r_b(\lambda)\left(\left[\sum_m i_m(i, j, t)\delta(\lambda - \lambda_m) + \ell_{\text{bkgd}}(\lambda, i, j, t)\right] * h(i,j)\right) d\lambda dt
\end{multline}
The summation over $n_{\text{frame}}$ image frames and the integration over time over the time-dependent terms can be replaced by a multiplication of $t_{\text{int}, b}$ (the integration time of the image taken with filter $b$) with the time-average of the terms that depend on $t$, which is denoted as $\bar{i}_m(i, j)$ and $\bar{\ell}_{\text{bkgd}}(\lambda, i, j)$. 
\begin{equation}
    \mathbb{E}\left[S_{\text{FOV}, b}(i, j)\right] = \frac{\Omega}{4\pi} t_{\text{int}, b} \int r_b(\lambda)\left(\left[\sum_m \bar{i}_m(i, j)\delta(\lambda - \lambda_m) + \bar{\ell}_{\text{bkgd}}(\lambda, i, j)\right] * h(i,j)\right) d\lambda
\end{equation}
The integration over $\lambda$ can be split into two pieces.
\begin{multline}
    \mathbb{E}\left[S_{\text{FOV}, b}(i, j)\right] = \frac{\Omega}{4\pi} t_{\text{int}, b} \int r_b(\lambda)\left[\sum_m \bar{i}_m(i, j)\delta(\lambda - \lambda_m)\right] * h(i, j)d\lambda \\ + \frac{\Omega}{4\pi} t_{\text{int}, b} \int r_b(\lambda) \bar{\ell}_{\text{bkgd}}(\lambda, i, j) * h(i, j) d\lambda
\end{multline}
The first term is an integral over the summation of Dirac delta distributions and can be simplified.
\begin{multline}
    \mathbb{E}\left[S_{\text{FOV}, b}(i, j)\right] = \sum_m \frac{\Omega}{4\pi} t_{\text{int}, b}r_b(\lambda_m) \bar{i}_m(i, j) * h(i, j)d\lambda \\ + \frac{\Omega}{4\pi} t_{\text{int}, b}\int r_b(\lambda) \bar{\ell}_{\text{bkgd}}(\lambda, i, j) * h(i, j) d\lambda
\end{multline}

The above equation is valid for each filter $b$. The four equations for the four filters can be stacked to yield a matrix equation:
\begin{multline}
    \begin{bmatrix}
        \mathbb{E}\left[S_{\text{FOV}, b_1}(i, j)\right] \\
        \mathbb{E}\left[S_{\text{FOV}, b_2}(i, j)\right] \\
        \mathbb{E}\left[S_{\text{FOV}, b_3}(i, j)\right] \\
        \mathbb{E}\left[S_{\text{FOV}, b_4}(i, j)\right]
    \end{bmatrix} = \frac{\Omega}{4\pi} \\ \begin{bmatrix}
        t_{\text{int}, b_1}r_{b_1}(1216) & t_{\text{int}, b_1}r_{b_1}(1304) & t_{\text{int}, b_1}r_{b_1}(1356) \\
        t_{\text{int}, b_2}r_{b_2}(1216) & t_{\text{int}, b_2}r_{b_2}(1304) & t_{\text{int}, b_2}r_{b_2}(1356) \\
        t_{\text{int}, b_3}r_{b_3}(1216) & t_{\text{int}, b_3}r_{b_3}(1304) & t_{\text{int}, b_3}r_{b_3}(1356) \\
        t_{\text{int}, b_4}r_{b_4}(1216) & t_{\text{int}, b_4}r_{b_4}(1304) & t_{\text{int}, b_4}r_{b_4}(1356) \\
    \end{bmatrix}\begin{bmatrix}
        \bar{i}_{\text{H1216}}(i,j) * h(i,j) \\
        \bar{i}_{\text{O1304}}(i,j) * h(i,j) \\
        \bar{i}_{\text{O1356}}(i,j) * h(i,j)
    \end{bmatrix} \\ + \frac{\Omega}{4\pi} \begin{bmatrix}
        \int r_{b_1}(\lambda) \bar{\ell}_{\text{bkgd}}(\lambda, i, j) * h(i, j) d\lambda \\
        \int r_{b_2}(\lambda) \bar{\ell}_{\text{bkgd}}(\lambda, i, j) * h(i, j) d\lambda \\
        \int r_{b_3}(\lambda) \bar{\ell}_{\text{bkgd}}(\lambda, i, j) * h(i, j) d\lambda \\
        \int r_{b_4}(\lambda) \bar{\ell}_{\text{bkgd}}(\lambda, i, j) * h(i, j) d\lambda
    \end{bmatrix}
\end{multline}

\begin{figure}[htbp]
    \centering
    \includegraphics[width=0.97\linewidth]{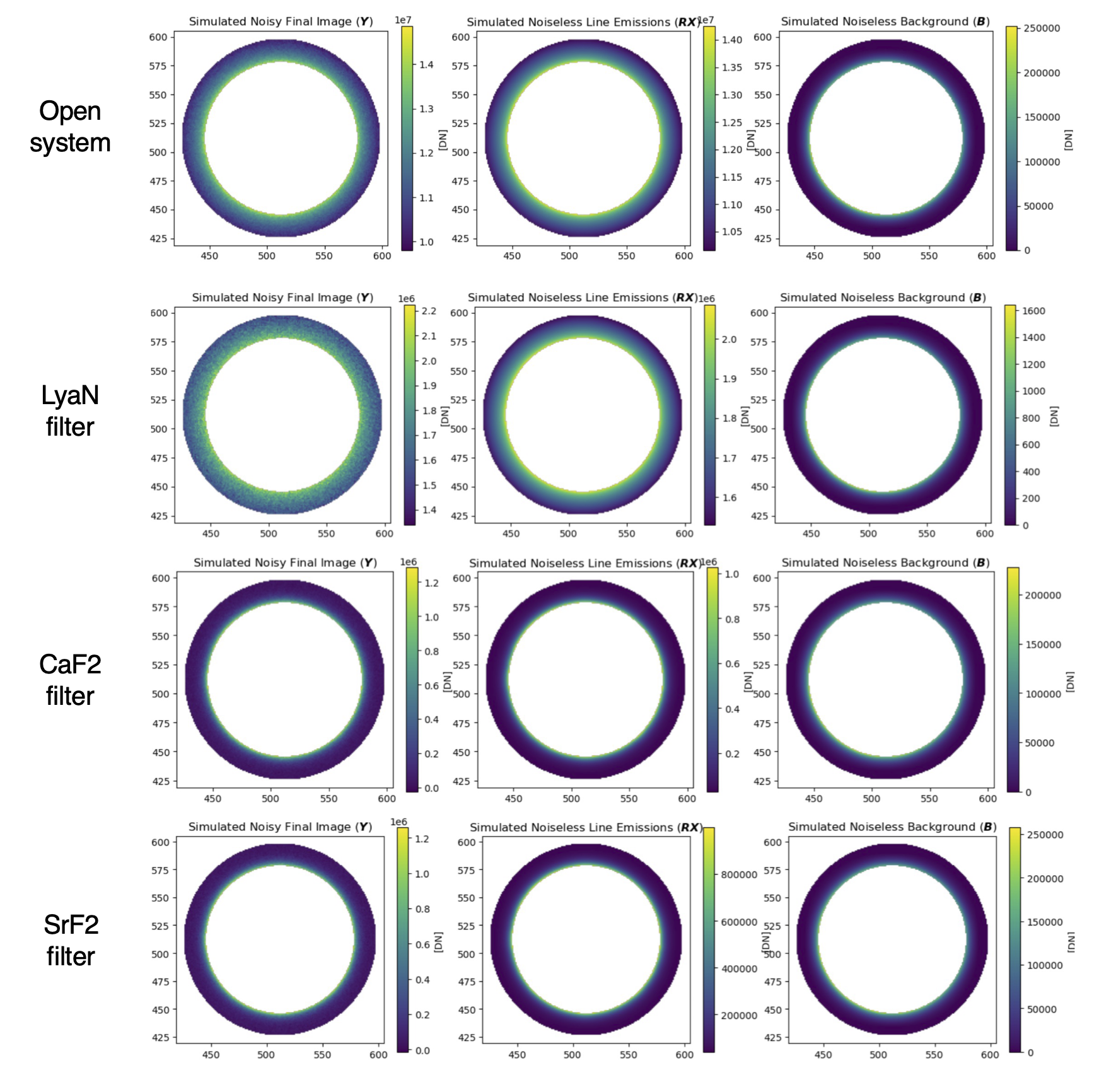}
    \caption{Simulated out-of-band signal components, restricted to pixels spanning the Earth's limb from $1.078 R_E < r_{ij} < 1.4 R_E$, for the open NFI system (top row), for the NFI narrowband LyaN filter (second row), for the NFI longpass CaF2 filter (third row), and the NFI longpass SrF2 filter (bottom row). These 2D spatial arrays are flattened into the row vectors in the following matrices in Equation \ref{eq:final_oob_eq_to_solve}: $\boldsymbol{Y}$ (left panels), $\frac{\Omega}{4\pi} \boldsymbol{RX}$ (center panels), and $\frac{\Omega}{4\pi} \boldsymbol{Z}$ (right panels).}
    \label{fig:oob_case_study}
\end{figure}

In the on-orbit operational environment, the theoretical expected value on the left-hand side is inaccessible; consequently, it is replaced by the stochastic measurement vector, denoted as $\vec{Y}(i, j)$. The right-hand side can be simplified by introducing some notation:
\begin{equation}
    \vec{Y}(i, j) = \frac{\Omega}{4\pi} \left(\boldsymbol{R}\vec{x}(i,j) + \vec{z}(i,j)\right)
\end{equation}
The above only describes a single pixel $i, j$. Equations for all the pixels can be combined into a single equation by adding more columns to $\vec{x}$, $\vec{z}$, and $\vec{Y}$, thus turning them into matrices.
\begin{equation}
\label{eq:final_oob_eq_to_solve}
    \boldsymbol{Y} = \frac{\Omega}{4\pi} \left(\boldsymbol{R}\boldsymbol{X} + \boldsymbol{Z}\right)
\end{equation}
Figures \ref{fig:oob_case_study} illustrates an example of the simulated morphological components of Equation \ref{eq:final_oob_eq_to_solve} across the four filter configurations. To construct the linear system, each 2D spatial array is flattened to populate the corresponding filter-specific row within the $\boldsymbol{Y}$, $\boldsymbol{RX}$, and $\boldsymbol{Z}$ matrices.

The algorithmic objective is to solve for $\boldsymbol{X}$ and $\boldsymbol{Z}$ utilizing the noisy measurement vector $\boldsymbol{Y}$ and the responsivity matrix $\boldsymbol{R}$ (derived via the responsivity retrieval algorithm detailed in Zhang et al. \cite{Zhang26c}). However, the inclusion of the $\boldsymbol{Z}$ term renders the system underdetermined, thereby precluding the use of standard least-squares minimization. Consequently, to reach a unique solution, it is necessary to impose specific physical constraints on $\boldsymbol{X}$ and $\boldsymbol{Z}$.

The structure that the solution $\boldsymbol{X}$ should take can be found by examining the original scene model for OOB photons found in Filippini et al. \cite{Filippini26}. Since the set of pixels $\boldsymbol{J}$ where OOB characterization is required is in the range $1.078< r_{ij} < 1.4$, which is completely within $1.055 < r_{ij} < 1.5$, $\boldsymbol{X}$ should take the form of a summation of three exponential decays as a function of radius, with one exponential decay for each line emission. Recall the mapping from pixel $i,j$ to pixel indices $p_{ij}, q_{ij}$, where the latter are the pixel indices of the closest in the set of pixels $\boldsymbol{D}$ that satisfy $r_{p_{ij}q_{ij}} < 1$ (see beginning of Section \ref{sec:oob_algo_head}). The algorithm therefore imposes a constraint where all channel pixels $i,j$ that map to the same pixel index $p_{ij}, q_{ij}$ are part of the same set of three exponential decays. Thus, each of these sets comes with six unknowns: three amplitudes $\beta_{m}[p_{ij}, q_{ij}]$ and three decay rates $\alpha_m[p_{ij}, q_{ij}]$. Unfortunately, radial exponentials are expected to change shape drastically under 2D spatial blurring by the instrument PSF. Thus, the images $\boldsymbol{Y}$ are first deblurred using the algorithm described in Section \ref{sec:deblur_algorithm}. Ideally, this removes all instances of $h(i,j)$ from Equation \ref{eq:final_oob_eq_to_solve} and increases modeling accuracy.

The background $\boldsymbol{Z}$ has two components: the Nitrogen Lyman-Birge-Hopfield (LBH) emission band spanning 1400-1800{\AA} \cite{lbh_definition}, which yields a relatively weak [DN] contribution due to the instrument's low responsivity at these wavelengths for all filters, and the Earth continuum albedo, which is only present due to instrument blur. Since images will be deblurred in order to add accurate constraint to $\boldsymbol{X}$, minimal error is incurred if the algorithm assumes that $\boldsymbol{Z}$ must follow the shape of a blurred Earth's disk that is then deblurred. Let this scaling factor be a column vector denoted $\vec{s}_{\text{bkgd}}(i, j)$. Thus, the $4\|\boldsymbol{J}\|$ unknowns in $\boldsymbol{Z}$ have been reduced to only $4$ unknowns: one amplitude $\beta_{\text{bkgd}, b}$ corresponding to the solar continuum albedo for each filter $b$. Let the four amplitudes be rewritten into a single vector $\vec{\beta}_{\text{bkgd}} = \begin{bmatrix}
    \beta_{\text{bkgd}, b_1}, & \cdots, & \beta_{\text{bkgd}, b_4} \\
\end{bmatrix}^T$. In the idealized theoretical limit with zero noise, a perfect instrument model, negligible LBH contributions, and a fully invertible deblurring operator, the background analytically simplifies to $\boldsymbol{Z} = \vec{\beta}_{\text{bkgd}} \vec{s}_{\text{bkgd}}^T$. Because the true physical background closely mirrors this fundamental shape, we adopt this reduced-rank structure as our operative constraint to solve the system. In practical on-orbit applications, deviations from these ideal conditions, such as residual blurring artifacts or nonzero LBH emissions, will manifest as structural approximation errors in the fitted background $\boldsymbol{Z}$. However, because this component constitutes a minor fraction of the total signal, the impact of these limitations is effectively bounded, as will be demonstrated by the end-to-end validation error analysis at the end of this section.

The final least-squares minimization is performed over the amplitudes and decay rates of all the line emission exponentials and the four unknown background amplitudes:
\begin{multline}
\label{eq:cost_function}
    \argmin_{\vec{\alpha}, \vec{\beta}, \vec{\beta}_{\text{bkgd}}} \left\| \text{deblur}\left(\boldsymbol{Y}\right) - \frac{\Omega}{4\pi}\left(\boldsymbol{R} \boldsymbol{X}' + \vec{\beta}_{\text{bkgd}} \vec{s}_{\text{bkgd}}^T \right)\right\|_2^2 \\\text{with } \boldsymbol{X}'(i, j) = \begin{bmatrix}
    \beta_{\text{H1216}}[p_{ij}, q_{ij}] \exp \left(\alpha_{\text{H1216}}[p_{ij}, q_{ij}] (r_{ij} - 1)\right) \\
    \beta_{\text{O1304}}[p_{ij}, q_{ij}] \exp \left(\alpha_{\text{O1304}}[p_{ij}, q_{ij}] (r_{ij} - 1)\right) \\
    \beta_{\text{O1356}}[p_{ij}, q_{ij}] \exp \left(\alpha_{\text{O1356}}[p_{ij}, q_{ij}] (r_{ij} - 1)\right) \\
    \end{bmatrix}
\end{multline}

The minimization is implemented numerically via Python's scipy.optimize library \cite{virtanen2020scipy}. Once $\vec{\alpha}$, $\vec{\beta}$, and $\vec{\beta}_{\text{bkgd}}$ are recovered, the final $\boldsymbol{X}$ and $\boldsymbol{Z}$ are reconstructed using
\begin{equation}
    \boldsymbol{X}(i, j) = \begin{bmatrix} \beta_{\text{H1216}}[p_{ij}, q_{ij}] \exp \left(\alpha_{\text{H1216}}[p_{ij}, q_{ij}] (r_{ij} - 1)\right) \\
\beta_{\text{O1304}}[p_{ij}, q_{ij}] \exp \left(\alpha_{\text{O1304}}[p_{ij}, q_{ij}] (r_{ij} - 1)\right) \\
\beta_{\text{O1356}}[p_{ij}, q_{ij}] \exp \left(\alpha_{\text{O1356}}[p_{ij}, q_{ij}] (r_{ij} - 1)\right) \\
\end{bmatrix}
\end{equation}
\begin{equation}
    \boldsymbol{Z} = \vec{\beta}_{\text{bkgd}} \vec{s}_{\text{bkgd}}^T
\end{equation}
The last two rows of $\boldsymbol{X}$ represent the O1304 and O1356 radiances in units of [photons/cm$^2$/s] in the altitude range 500-1500km, while the four rows of $\boldsymbol{Z}$ represent the [DN] contribution for images taken by the four filters from all OOB photons that are not the three line emissions.

The OOB removal algorithm proceeds by scaling the signal components derived by the retrieval algorithm by the integration time of the input image, followed by subtraction. This correction is applied exclusively to the NFI pixel region of interest.

The OOB algorithm is validated by using the numerical image simulator to generate a sequence of four images on the NFI channel using four filters (open, LyaN, CaF$_2$, and SrF$_2$) with integration times of (30min, 60min, 30min, and 30min) respectively, followed by a 30-minute storm-time (S3) LyaN image on the NFI channel. These settings correspond to a realistic, pre-flight beginning-of-life case study. Simulated L0 images were processed through the beginning of the calibration pipeline (instrument effect correction, star/Moon/outer planet masking and interpolation, IPH removal, and deblurring), with supporting images generated using the numerical image simulator as necessary. These images were then processed by the OOB retrieval algorithm described in this section.

\begin{figure}[htbp]
    \centering
    \includegraphics[width=0.5\linewidth]{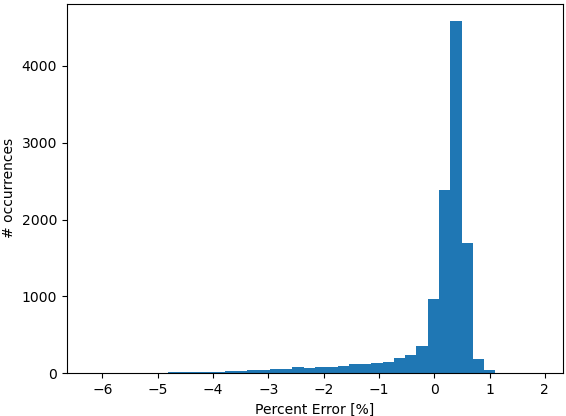}
    \caption{Histogram of the Out-of-Band (OOB) photon characterization error. The reported percent error represents the absolute OOB retrieval error normalized by the target exospheric Ly-$\alpha$ signal, evaluated over NFI channel pixels within the spatial region of interest ($1.078R_E < r_{ij} < 1.428R_E$).}
    \label{fig:oob_ret_val}
\end{figure}

This analysis focuses exclusively on the NFI channel within the operational region of interest ($1.078R_E < r_{ij} < 1.428R_E$), where OOB contributions must be actively mitigated. Because the OOB intensity in this region is typically a minor fraction of the total signal, algorithm performance is validated relative to the target Ly-$\alpha$ emission. Specifically, a larger relative error within the faint OOB component is acceptable provided the error's absolute magnitude remains negligible compared to the primary Ly-$\alpha$ signal. Percent error statistics are therefore calculated as $100 \times $ (error-in-OOB) / (Ly-$\alpha$-signal). As illustrated in Figure \ref{fig:oob_ret_val}, the OOB characterization error, normalized by the exospheric Ly-$\alpha$ signal, yields a mean bias of $0.03\%$ and a standard deviation of $0.88\%$. The science requirements traceability matrix mandates a maximum allowable error of $20\%$ relative to the Ly-$\alpha$ signal for baseline mission success \cite{Waldrop26}. With both the mean and standard deviation falling well below this threshold, the proposed OOB background estimation algorithm successfully satisfies the mission requirement.

\section{Science Data Processing Pipeline: L1B to L1C}
\label{sec:sci_data_processing_l1B_l1C}

This section summarizes the calibration steps in the L1B$\rightarrow$L1C data processing pipeline, which ingests L1B science images, i.e., instrument effect corrected images of Earth nadir acquired using the narrowband Lyman-$\alpha$ filter LyaN.  For more details on how L1B images are derived from their parent L1A images, see Zhang et al. \cite{Zhang26a}.  The L1B$\rightarrow$L1C processing pipeline removes photon backgrounds in the nadir scene, based on estimation of background signals from off-nadir and out-of-band filter calibration data as described in the sections above, and converts the isolated exospheric Lyman-$\alpha$ signals into physical units of absolute emission radiance.   Absolute sensitivity calibration requires knowledge of the optical system responsivity at Ly-$\alpha$ (in [DN/sec] per unit radiance), which is derived on-orbit from measurements of known stellar sources (see Zhang et al. \cite{Zhang26c} for details on stellar target selection and the responsivity retrieval algorithm formulation).  As described in that work, performance testing of the responsivity retrieval algorithm using numerical simulations of stellar data \cite{Filippini26} 
demonstrated that the stellar calibration approach is capable of high fidelity retrieval of the narrowband science channel responsivity at Lyman-$\alpha$. Specifically, that analysis concluded that that an ensemble of 30 distinct stellar targets supports retrieval to an accuracy of $0.11\% \pm 3.81\%$ for the NFI LyaN configuration and $-0.09\% \pm 4.15\%$ for the WFI LyaN configuration \cite{Zhang26c}.  

The steps of the L1B$\rightarrow$L1C science data processing algorithm are summarized as follows:
\begin{enumerate}
    \item Retrieve the locations of all stars in the image using the algorithm described in Section \ref{sec:celestial_body_alg}.
    \item Retrieve the locations of the lunar disk and those of the outer planets using the algorithm described in Section \ref{sec:moon_and_planet_loc}.
    \item Mask all pixels that are within a full PSF width of the identified celestial objects. Interpolate over smaller holes using the algorithm found in Section \ref{sec:hole_interp}.
    \item Retrieve the cross-channel Ly-$\alpha$ responsivity ratio using the algorithm described in Section \ref{sec:cross_channel_responsivity}.
    \item Retrieve and subtract the mean [DN] contribution from IPH photons in the image using the algorithm found in Section \ref{sec:iph_retrieval_algorithm}.
    \item Deblur the image using the algorithm discussed in Section \ref{sec:deblur_algorithm}.
    \item If the image is from the NFI channel, subtract the most recent mean [DN] contribution from OOB photons from the calibration database.
    \item Divide by the absolute Ly-$\alpha$ responsivity to yield the L1C image in physical units of $\left[\frac{\text{ photons}}{\text{cm}^2 \cdot \text{s}} \right]$.
\end{enumerate}

The result of these operations are denoted L1C images, which are reported as Lyman-$\alpha$ radiance in units of [photons/s/cm$^2$]. To assess the performance of the end-to-end L0 to L1C calibration pipeline, the numerical image simulator was utilized to generate synthetic L0 data for both channels corresponding to a realistic, beginning-of-life case study \cite{Filippini26}.  The scene simulation featured exospheric Lyman-$\alpha$ radiance computed from an assumed static H density distribution, modeled IPH background, modeled out-of-band emission from Earth's limb and disk, and expected celestial sources in the FOV of both channels based on the realistically-specified vantage and viewing geometry. The simulated observing configuration employed the baseline science filter (LyaN) and adopted integration times of $30$ minutes for NFI and $60$ minutes for WFI, designed to mimic the nominal parameters of planned nadir-pointed science observations. Only a single set of simulated science images, corresponding to a single noisy image acquision for each channel from a  fixed on-orbit vantage, was produced for this test, owing to the computational expense of both scene and image generation.  However, the test was supported by a complete synthetic set of baseline, non-science calibration images (such as the out-of-band, IPH mapping, and flat-field sequences). 

End-to-end L0$\rightarrow$L1C pipeline performance using the simulated images was assessed in terms of the residual error statistics between the retrieved exospheric Lyman-$\alpha$ emission radiances across the FOVs of the respective channels and the ground-truth radiances that were ingested into the numerical simulator to produce the input L0 images.  Across the non-disk portion of the NFI FOV, the mean percent error of the retrieved exospheric radiance relative to the ground-truth is $-2.1\%$, with a standard deviations of $12.4\%$. For the WFI channel, the absolute mean percent error is $-2.9\%$, with a standard deviation of $9.0\%$.

While the deblurring algorithm introduces increased noise variance at the L1C stage relative to L1B, the overall accuracy remains exceptional. The science requirements traceability matrix mandates an overall calibrated radiance accuracy of $30\%$ and precision of $20\%$ to achieve baseline mission success \cite{Waldrop26}. Because the evaluated mean bias and standard deviation both fall well below this threshold, the proposed science data processing pipeline satisfies this operational requirement with significant margin.

\section{Conclusion}

The Carruthers Geocorona Observatory successfully launched in September 2025 and began science operations at the Earth-Sun Lagrange 1 (L1) point in March 2026. This work presents the end-to-end science data processing pipeline required to transform instrument-corrected images into absolutely calibrated exospheric H Lyman-$\alpha$ radiance. Validation via numerical simulation demonstrates that the baseline concept of operations governing calibration and science image acquisition, along with the algorithms developed to estimate and remove photon backgrounds from the science images, meet mission measurement requirements with significant margin in support of the exospheric H density retrieval algorithms detailed in the accompanying papers of this issue.

\subsection*{Acknowledgements}

\begin{itemize}
\item Funding: This work was supported by the NASA Science Mission Directorate, Heliophysics Division through award 80GSFC21C0038.
\item Conflict of interest/Competing interests:
Not applicable
\item Ethics approval and consent to participate:
Not applicable
\item Consent for publication:
Not applicable
\item Author contribution: A. Zhang led the conceptualization and algorithm development for Sections 4-6 (except Section 5.1), led the validations for Sections 3-6, and led the manuscript writing. H. Filippini led the conceptualization and algorithm development for Section 3. J. Craig led the conceptualization and algorithm development for Section 5.1.  As the Principle Investigator of the mission, L. Waldrop provided project supervision, conceptualization, and funding acquisition, as well as assistance in the preparation of this manuscript. J. Clarke, F. Kamalabadi, and P. Joshi provided methodological supervision.
\item AI use: During the preparation of this work, the authors used Gemini to improve the readability and language of the manuscript and assist in debugging Python code. After using this tool, the authors reviewed and edited the content and code as needed and take full responsibility for the integrity and content of the publication.
\end{itemize}

\bibliographystyle{ieeetr} 
\bibliography{ref}

\end{document}